\documentclass[preprints,article,accept,moreauthors,pdftex]{Definitions/mdpi}

\firstpage{1} 
\pubvolume{1}
\issuenum{1}
\articlenumber{0}
\pubyear{2022}
\copyrightyear{2022}
\datereceived{} 
\dateaccepted{} 
\datepublished{} 
\hreflink{https://doi.org/} 

\Title{SIDIS Measurement with A=3 Nuclei}

\TitleCitation{SIDIS Experiments with A=3 Nuclei}

\Author{Zhihong Ye $^{1}$}

\longauthorlist{yes}
\Author{Zhihong Ye$^{1}$*, Dipangkar Dutta$^{2}$, Dave Gaskell$^{3}$, Or Hen$^{4}$, Dave Meekins$^{3}$, Dien Nguyen$^{3}$, Jennifer Rittenhouse West$^{5}$ and Lawrence Weinstein$^{6}$}

\AuthorNames{Zhihong Ye, Dipangkar Dutta, Dave Gaskell, Or Hen, Dave Meekins, Dien Nguyen, Jennifer Rittenhouse West and Lawrence Weinstein}

\AuthorCitation{Ye, Z.H.}

\AuthorCitation{Ye, Z. H.; Dutta, D.; Gaskell, D.; Hen, O.; Meekins, D.; Nguyen, D.; Rittenhouse West, J.; Weinstein, L.}

\address{%
$^{1}$ \quad Department of Physics, Tsinghua University, Beijing, 100084 China\\ 
$^{2}$ \quad Mississippi State University, Mississippi State, MS 39762 USA \\
$^{3}$ \quad Thomas Jefferson National Accelerator Facility, Newport News, Virginia 23606, USA\\
$^{4}$ \quad Massachusetts Institute of Technology, Cambridge, Massachusetts 02139, USA\\
$^{5}$ \quad Lawrence Berkeley National Laboratory, Berkeley, CA 94720, USA\\
$^{6}$ \quad Old Dominion University, Norfolk, Virginia 23529, USA}

\corres{Correspondence: yez@tsinghua.edu.cn}

\abstract{
We introduce a new experimental effort at Jefferson Lab (JLab) to precisely measure the ratios of charged pion electroproduction in Semi-Inclusive Deep Inelastic Scattering (SIDIS) from $^2$D, $^3$He, and $^3$H targets~\cite{c12-21-004}. This conditionally approved experiment (C12-21-004) aims to run in Hall-B using the standard CLAS12 configuration and a new target system developed for the approved quasi-elastic experiment (E12-20-005). In this data-driven study, we will measure the cross-sections as a function of ($x$, $Q^2$, $z$, $P_T$) to allow the extraction of the unpolarized parton distribution functions (PDFs), transverse momentum distributions (TMDs) and fragmentation functions (FFs) in A = 3 nuclei. By using super-ratios of pion yields of SIDIS off light nuclei over a wide $x_B$ range, we search for evidence of a flavor dependence in the EMC effect, giving us new insights into the effect of the nuclear environment on valance quarks. Double-ratios between A = 3 mirror nuclei can provide a direct measurement of the d/u ratios at large x due to their similar and well-understood nuclear corrections. With the utilization of mirror nuclei and the large kinematic range, and high precision of CLAS12, we will be able to maintain the sensitivity to the underlying physics but dramatically decrease the nuclear uncertainties due to attenuation and hadronization in heavy nuclei targets.
}

\keyword{Tritium; SIDIS; Nuclear Medium Modification; EMC Effect} 

\begin{document}
\maketitle
\section{Introduction}\label{intro}
Nuclei are stable systems made of quarks and gluons bound together by the strong force described in the theory of Quantum Chromodynamics (QCD). A full understanding of the confining nature of QCD requires exploring the mechanism of quarks and gluons interaction via strong force in nuclear environment \cite{Brambilla:2014jmp,Shuryak:1980tp}. Effective theories provide a good description of the nuclear dynamics by treating nucleon-nucleon interaction via exchanging mesons in average distances, but a deeper investigation is needed to understand the involvement of quarks and gluons in forming nuclei. Over the last several decades a significant amount of effort has been invested in the study of nuclear structure at small distances to map out the quark distributions inside nuclei and find novel signatures of QCD in nuclei. 

One of the most important discoveries is the so-called ``EMC effect''~\cite{Aubert83} which revealed that nuclei are not simply collections of free nucleons undergoing Fermi motion in the nucleus but the nuclear environment affects quark distributions in bound nucleons in an unexpected way. Since this effect was discovered by the EMC collaboration in 1983, enormous theoretical and experimental efforts have been invested to understand the modification of the structure functions in nuclei and their origin \cite{Geesaman95,Norton03,Malace2014,Hen:2016kwk}. Experiments at SLAC E139 ~\cite{Gomez94} and JLab~\cite{Seely09} precisely mapped out the $x$-dependence of the $A/D$ cross-section ratios by measuring the inclusive DIS electron scattering off light to heavy nuclei. However, none of the inclusive measurements distinguished between the contributions of individual quarks to the modification of the nuclear structure nor did they provide sufficient insight into the underlying QCD mechanism. Many theoretical approaches, such as nuclear binding and convolution, dynamical quark rescaling, quark-meson coupling~\cite{Cloet06}, and more, have attempted to describe the experimental observations of the EMC effect. However, although most of the models can describe some aspects of the data well, none of them is a generally accepted description of the EMC effect nor touches on the origins of distorted quark behavior. 

An encouraging discovery in recent years is the experimentally verified connection between the EMC effect and short-range correlations (SRC)~\cite{weinstein11,Fomin2011ng,Hen12,Hen:2016kwk}. They discovered a linear correlation between the number of SRC pairs and the size of the EMC effect, suggesting two leading hypotheses of for its origin:
\begin{itemize}
    \item The EMC effect is flavor-independent due to a universal mechanism for all nucleons that form a high-density cluster in nuclei;
   \item  The EMC effect is flavor-dependent in asymmetric nuclei and could be a direct result of modification of the SRC pairs in nuclei~\cite{Schmookler:2019nvf}.
\end{itemize}
The second hypothesis appears to be preferred after careful examination of the exclusive SRC experimental data ~\cite{Cruz-Torres:2019fum}. New theoretical studies of diquark degrees of freedom in $A=3$ \cite{West:2020tyo} and $A=4$ \cite{West:2020rlk} nuclei also predict isospin-dependent SRCs. Other simple descriptions also can naturally produce larger effects in protons compared to neutrons in asymmetric nuclei~\cite{cloetpc,Malace2014}. Moreover, the first indication of an isospin dependent EMC effect was recently reported by combining the $^3$He/$^3$H data from the MARATHON experiment~\cite{Adams21} with a global QCD analysis to simultaneously extract PDFs and nuclear effects in $A$ = 2 and $A$ = 3 nuclei~\cite{Cocuzza:2021rfn}. New experimental data with more sensitive techniques are required to obtain unambiguous evidence of the flavor dependence of the EMC effect, providing insight into the origin of the EMC effect and its connection to SRCs. 

Semi–inclusive DIS (SIDIS) can be used as a “flavor tag” to look for signatures of differences in the EMC effect in the $u$- and $d$- quark distributions in nuclei. In particular, measurements of SIDIS on mirror nuclei is one of the most effective ways to enhance the sensitivity of this technique, while simultaneously reducing the impact of hadron attenuation. The SIDIS measurement on $A=3$ mirror nuclei will provide direct access (at leading order) to the nucleon $d/u$ ratio at high $x$ because the measurement is free of most theoretical corrections. Precise data will also enable us to explore the difference of the nuclear effect in isotopes unseen by inclusive measurements. With its entirely different systematics, the SIDIS measurement will be complementary to the recent DIS measurements by MARATHON~\cite{MARATHON} and the spectator tagged measurements by BoNuS12~\cite{bonus12}, BAND and LAD experiments~\cite{BANDandLAD}. The SIDIS measurement on $A=3$ nuclei will also provide the first high precision determination of the fragmentation functions (FF) and unpolarized transverse momentum dependent distributions (TMD), which are important for establishing the 3D structure of nucleons, a major priority for a complete QCD-based understanding of the nucleon. 

With the golden opportunity of a tritium target that is once again (likely for the last time) approved to be used at JLab~\cite{Hen:2020wxa}, we propose to measure the absolute cross-sections of the SIDIS $\pi^{\pm}$ production with 10.6 Gev electrons scattering off $^2$D, $^3$H, and $^3$He~\cite{c12-21-004}.  The wide acceptance of CLAS12 allows us to perform multi-dimensional binning (e.g.,  $Q^2$, $x$, $z$, $P_T$) of the data to help disentangle the $x$- and $z$-dependencies. These will enable theorists to study with great detail medium effects on the intrinsic quark distributions and the quark hadronization in the different nuclei. We will probe the flavor-dependent EMC effect in $A=3$ nuclei, obtain the $d/u$ ratios at high-x by taking advantage of the theoretically well-controlled nuclear effects in $^3$H and $^3$He, and map out the three-dimensional parton distributions functions and fragmentation functions. The high-precision SIDIS data will ultimately allow us to study the factorization theorem and understand the hadronization process in light nuclei. It is essentially a data-driven experiment aimed at providing high-quality data for theorists to perform sophisticated global analysis.

\subsection{Experimental Setup}
\begin{figure}[H]
\centering
\includegraphics[width=0.8\textwidth]{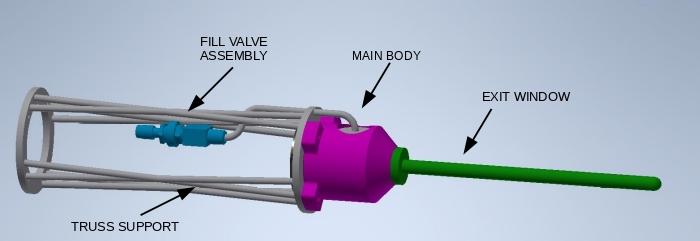}
\caption{The conceptual design of the new gas target cell to be used in E12-12-005 and this experiment.}
\label{T2-cell} 
\end{figure}

We will use the standard configuration CLAS12 detector and an unpolarized beam incident on identical $^2$D, $^3$H, and $^3$He target cells to measure $(e,e'\pi^+)$ and $(e,e'\pi^-)$. Shown in Fig.~\ref{T2-cell}, a new target system designed for another newly approved Tritium-SRC experiment (E12-12-005~\cite{Hen:2020wxa}) will be used to contain these gases. The luminosity of these gas targets will be controlled at $3.45 \times 10^{34}~nucleons\cdot cm^{-2}\cdot s^{-1}$ to accommodate the DAQ limit. The experiment will extract ratios of sum and difference between $\pi^\pm$ events which have different acceptance in CLAS12, so we require the polarity of the CLAS12 torus to be regularly flipped to minimize the systematic uncertainties. We requested a beam time of 10 PAC days for $^2$D and 20 PAC days for each $A=3$ target, plus additional 10 days of configuration changes and calibration runs. 

\section{Experimental Goals}
\subsection{3D Distributions in $A=3$}
One of the key physics programs in the JLab 12GeV era and the future Electron-Ion Collider (EIC) era is to map out the three-dimensional structure of the nucleon and understand how the spin and orbital angular momentum of individual quarks contribute to the nucleon's spin. Using electrons scattering off protons, $^2$D and $^3$He polarized longitudinally or transversely, one can measure different asymmetries in the SIDIS reaction with additional detection of the transverse momenta of the outgoing hadrons. These asymmetries allow the extraction of the Transverse Momentum Distributions (TMDs) after being combined with corresponding unpolarized cross-sections which require good knowledge of unpolarized TMDs and unpolarized Fragmentation Functions (FFs) as inputs. In particular, with the great experimental precision to be achieved at JLab and EIC, understanding the nuclear effect in SIDIS data using $^2$D and $^3$He as effective neutrons will soon become as important as those in inclusive DIS data which are still being intensively studied.

At LO the unpolarized SIDIS cross-section is expressed as:
\begin{equation}
    \frac{d\sigma_{UU}^h}{dxdQ^2dzdP_T} \propto \sum_{q}e^2_q[f^q_1(x, k_{\perp})\otimes D_1^{q\rightarrow h} (z, q_{T})],
\end{equation}
where $f^q_1(x, k_{\perp})$ and $D_1^{q\rightarrow h} (z, q_{T})$ are the 3D unpolarized TMDs and FFs. $P_T$ is the transverse momentum of the detected hadron, $k_{\perp}$ is the initial transverse momentum of the quark before being knocked out, and $q_{T}$ is the final transverse momentum of the struck quark before hadronization happens. At LO approximation, there is a relation among three quantities, $\vec{P}_{T} = z\Vec{k}_{\perp} + \vec{q}_{T} + O(k_{\perp}^2/Q^2)$. The exact forms of the $k_{\perp}$ distributions in TMDs and the $q_{T}$ distributions in FFs are largely unknown due to limited experimental precision. A so-called Gaussian-Ansetz has been widely used in the global analysis~\cite{Anselmino:2013lza} of the SIDIS data to extract distribution for the transverse component of both the TMDs and FFs. However, this approximation is largely oversimplified. For example, the Gaussian distribution is not necessarily the same for all quark-flavors; Some analyses show the Gaussian-width ($<P_T^2>$) also has sizable $x$-dependence~\cite{Matevosyan:2011vj} and even observe the non-Gaussian behavior~\cite{Nematollahi}.  

Highly quality 4D ($Q^2$,$x$,$z$,$P_T$) SIDIS data at a wide kinematic range are essential to map out the $P_T$ distribution for individual quarks. They can further allow us to decouple the $f^q_1(Q^2, x,k_{\perp})$ from $D_1^q(Q^2, x,q_{T})$ in a detailed global analysis with helps from theoretical models. One also can further study the azimuthal distribution of the unpolarized SIDIS events ($\phi$) which provides precious information on the Boer-Mulder TMD as well as the high-order and high-twist QCD effects~\cite{Barone:2009hw}. 

\begin{figure}[ht]
    \centering
    \includegraphics[width = 1.10\textwidth, angle =0]{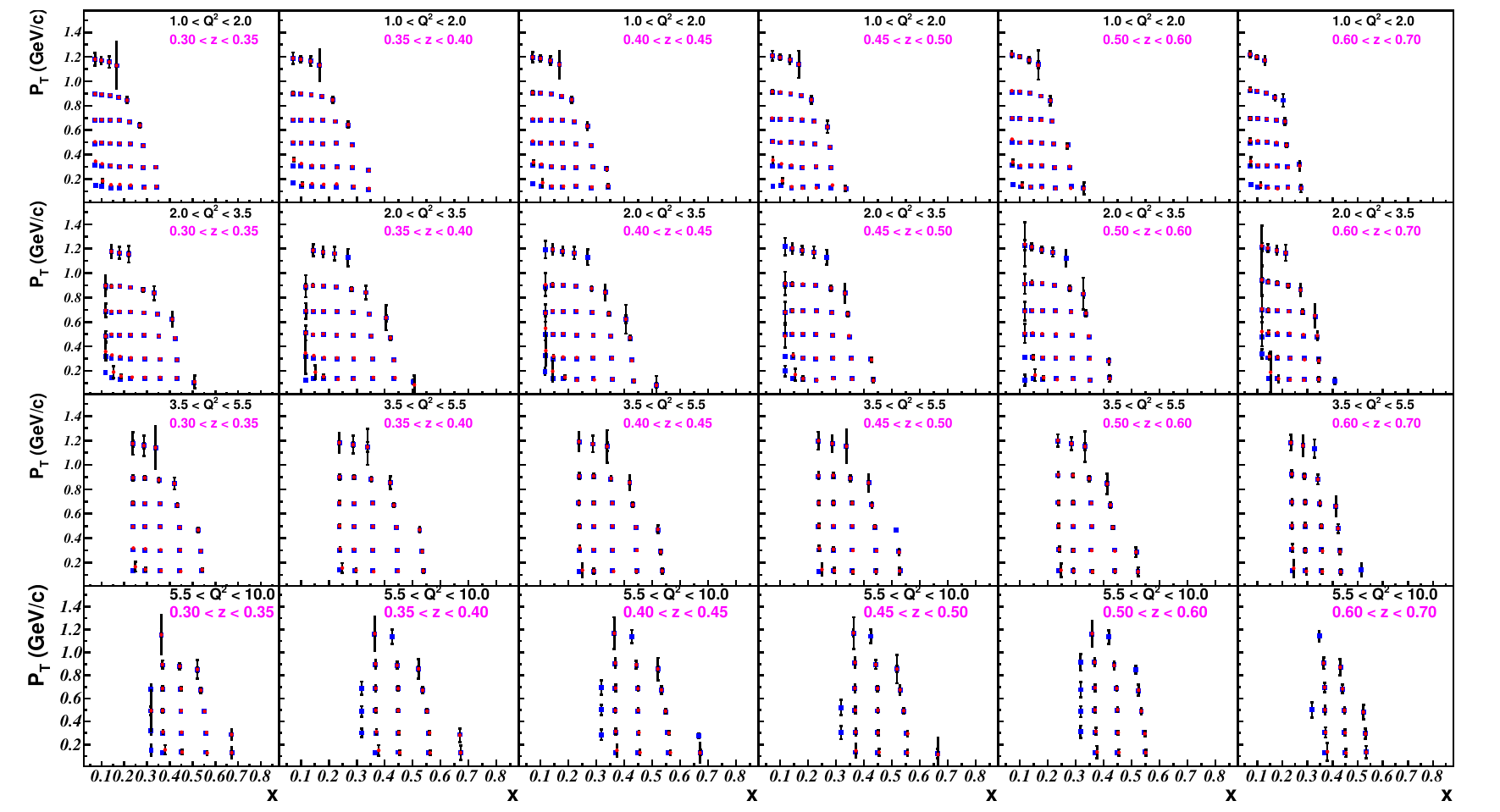}
    \caption{Projected $^3$H Pion-SIDIS data binned in 4D. Blue (red) dots are $\pi^+$ ($\pi^-$). SIDIS events were selected by applying the standard cuts: $Q^{2}>1$ (GeV/c)$^{2}$, $W^2 > 4$ (GeV/c)$^2$, $0.1 < y < 0.85$, and $0.3<z < 0.8$.$^2$D and $^3$He have similar projected coverage and precision.}
    \label{h3_pion_4d}
\end{figure}

In this experiment, we will provide high precision SIDIS data in 4D ($Q2$, $x$, $z$, $P_T$) as shown in Fig.~\ref{h3_pion_4d}. The measurements are based on the ratios of pion yields ($\pi^+ \pm \pi^-$) from two different targets, in which many systematic uncertainties largely cancel. Identical target cells will be used for the different target gases. To minimize the uncertainties of the $\pi^{\pm}$ acceptances we will flip the Torus polarity frequently. Combining other standard CLAS12 calibration and analysis techniques, we will control random coincidence background below 0.5\%, the point-to-point systematic uncertainties in the ratios to better than 1\% (see Table~\ref{unc}), and the normalization uncertainties in the ratio of about 2.5\%. We also include the uncertainty from the beam-charge measurement run-by-run stability (1\%), as well as the tritium decay correction (0.15\%~\cite{Cruz-Torres:2020uke}).
\begin{table}[ht!]
    \centering
    \begin{tabular}{c|c|c|c|c|c}
         & Sectors & Tracking & Vertex & Fiducial & Acceptance  \\
         \hline
       Uncertainty (\%)  & 0.34 & 0.13 & 0.16 & 0.41 &0.1  \\
       \hline
    \end{tabular}
    \caption{The systematic uncertainties of the $(e^+,e^{+'} p)/(e^-,e^{-'} p)$ ratio  in the CLAS Two-Photon Exchange experiment \cite{Rimal:2016toz}.  ``Sectors'' refers to the 6-sector ratio variance, ``tracking'' refers to the CLAS charge-dependent tracking, ``vertex'' refers to the effect of different vertex cuts, ``fiducial'' refers to the effect of narrowing the fiducial cuts, and ``acceptance'' refers to the maximum variance in the acceptance correction factors determined from monte carlo simulation.  Since the TPE systematic uncertainties were determined by statistical measurements, we used the data bins with the greatest statistical precision for this table.}
    \label{unc}
\end{table}

With $A=3$ isospin symmetric nuclei in which nuclear effects are small and under control, we can carefully examine the $P_T$ distributions of different quark flavors and examine nuclear effects in comparison to free nucleons. These data will provide enormous value to the TMD community for the study of the factorization theorem, the hadronization process, and most importantly, the global analysis of exacting unpolarized TMDs and FFs. The precise determination of their TMDs and FFs, combined with good theoretical control to minimize the nuclear effect, can be directly applied to the studies of other TMDs measured by polarized proton and $^3$He (as effective neutrons) targets where unpolarized results are used as inputs.

\subsection{Flavor-Dependent EMC Effect}
\begin{figure}[ht!]
    \centering
     \includegraphics[width = 1.0\linewidth]{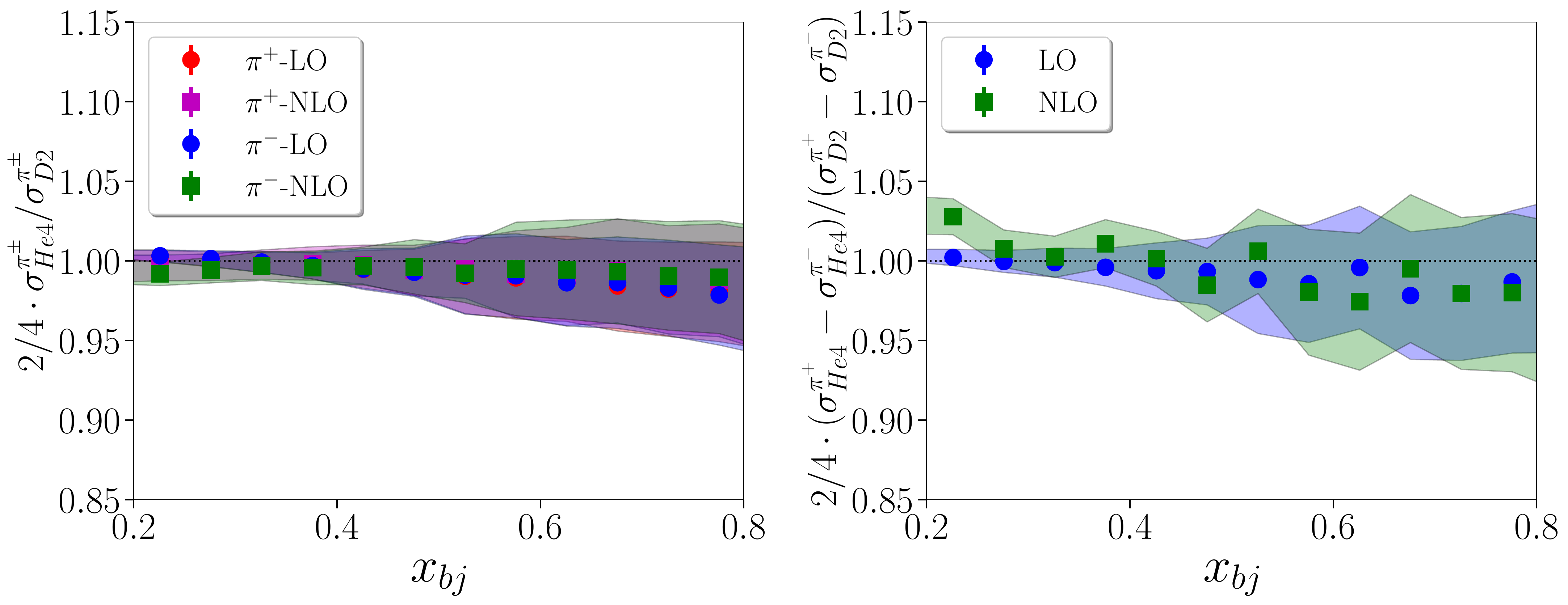}
    \caption{ $\sigma^{\pi{^\pm}}_{He4}/\sigma^{\pi{^\pm}}_{D2}$ (left) and $R^{\pi,-}_{He4/D2}$ (right)  calculated with nNNPDF1.0~\cite{Ball:2008by} and NNFF1.0~\cite{Bertone:2017tyb} for both LO and NLO cases. The bands are the statistical errors from the global fit.}    
    \label{nlo_lo_comp}
\end{figure}

After integrating over $P_T$, the LO SIDIS cross-section for an unpolarized nucleus reduces to:
   \begin{align}
    \frac{d\sigma_A^h}{dxdQ^2dz}=\frac{4\pi \alpha^2s}{Q^4}(1-y+\frac{y^2}{2}) \cdot A \cdot \sum_{q}e^2_q f_A^q(x)\cdot D_A^{q\rightarrow h}(z), \label{eq_sf_dis_sidis},
  \end{align}
where $f^q_A(x)$ is the average unpolarized parton distribution functions (PDFs) of the same quark-flavor in all protons and neutrons. Under isospin-symmetry, we have:
\begin{equation}
    u_A =\frac{Z\tilde{u}_{p,A} + N\tilde{u}_{n,A}}{A}=\frac{Z\tilde{u} + N\tilde{d}}{A},~~~d_A = \frac{Z\tilde{d}_{p,A} + N\tilde{d}_{n,A}}{A}=\frac{Z\tilde{d} + N\tilde{u}}{A}. \label{eq_npdf_pA}
\end{equation}
Meanwhile, $D_A^{q\rightarrow h}(z)$ is unpolarized 1D Fragmentation functions (FFs) in nucleus-A, respectively. Due to the strong correlations of the struck quark and the leading hadron in SIDIS,  a common approach in the global analysis of the SIDIS nucleon data is to simplify the FFs into a set of ``favored" (labeled as "+" sign) and ``unfavored" (labeled as "-" sign) terms,e.g., $D_u^{\pi^+} =D_{\bar{d}}^{\pi^+} = D_d^{\pi^-}=D_{\bar{u}}^{\pi^-} \equiv D^{+}$, $D_d^{\pi^+} =D_{\bar{u}}^{\pi^+} =D_u^{\pi^-}=D_{\bar{d}}^{\pi^-} \equiv D^{-}$, and $D_s^{\pi^{\pm}} = D_{\bar{s}}^{\pi^{\pm}} = D^s$. We adopt a similar approach to the nuclear FFs. After ignoring heavy quarks we can rewrite the per-nucleon SIDIS cross-sections for a nucleus as: 
\begin{equation}
\begin{aligned}
 		\sigma_A^{\pi^{\pm}}(x,z) /A &\propto e_u^2 \cdot {\color{black} u_A(x)} \cdot {\color{black} D_A^{\pm}(z)} \hspace{0.0cm} + e_u^2 \cdot {\color{black}\bar{u}_A(x)} \cdot{\color{black}D_A^{\mp}(z)} \\
 		&+ e_d^2 \cdot {\color{black}d_A(x)} \cdot {\color{black}D_A^{\mp}(z)} \hspace{0.01cm}+ e_d^2 \cdot {\color{black}\bar{d}_A(x)} \cdot {\color{black}D_A^{\pm}(z)} \\
 		&+ e_s^2 \cdot {\color{black}s_A(x)} \cdot {\color{black}D_A^{s}(z)} \hspace{0.06cm}+ e_s^2 \cdot {\color{black}\bar{s}_A(x)} \cdot {\color{black}D_A^{s}(z)}, \\
\end{aligned}
\end{equation}
where $e_u = 2/3$, $e_d=-1/3$, and $e_s=-1/3$. One can then construct the charge-sum and charge-difference of the SIDIS data:
\begin{align}
  (\sigma_A^{\pi^+}  \pm  \sigma_A^{\pi^-})/A &\propto {\color{black}[4(u_A\pm  \bar{u}_A) \pm (d_A\pm  \bar{d}_A)]}\cdot{\color{black}[D_A^{+} \pm  D_A^{-}]} \label{A_ypm},
\end{align}
where the strangeness part  completely cancels in the charge-difference and is neglected in the charge-sum since we focus on $x>0.1$. We can take the ratios of Eq,~\eqref{A_ypm} between two nuclei:
\begin{equation}
\begin{aligned}
    R^{\pi,\pm}_{A_1/A_2}(x,z) & \simeq \frac{(\sigma_{A_1}^{\pi^+}  \pm  \sigma_{A_1}^{\pi^-})/A_{1} }{(\sigma_{A_2}^{\pi^+}  \pm  \sigma_{A_2}^{\pi^-})/A_{2} }\simeq {\color{black}\frac{4(u_{A_1} \pm \bar{u}_{A_1} ) \pm (d_{A_1}  \pm \bar{d}_{A_1} )}{4(u_{A_2} \pm \bar{u}_{A_2} ) \pm (d_{A_2}  \pm \bar{d}_{A_2} )}} \cdot B^{\pm}_{A_{1}/A_{2}}(z),
    \label{eq_rpm} 
    \end{aligned}
\end{equation}
where $B^{\pm}_{A_{1}/A_{2}}(z)=(D_{A_1}^{+}\pm D_{A_1}^{-})/(D_{A_2}^{+}\pm D_{A_2}^{-})$. Calculation of Eq~\eqref{eq_rpm} for $^4$He with the SIDIS cross-sections at the next-leading-order (NLO)  suggests a negligible difference compared with the LO approach which is presumably similar for lighter nuclei as shown in Fig~\ref{nlo_lo_comp}.

$R^{\pi,+}_{A/D}(x,z)$ directly links to the EMC ratio, $R_{EMC}$, obtained from the inclusive measurement when ignoring heavier quarks, i.e.:
\begin{equation}
  R^A_{EMC} =\frac{2}{A}\frac{\sigma^{DIS}_A(x)}{\sigma^{DIS}_D(x)} \simeq {\color{black}\frac{4(u_{A} + \bar{u}_{A} ) + (d_{A}  + \bar{d}_{A} )}{4(u_{D} + \bar{u}_{D} ) + (d_{D}  + \bar{d}_{D} )}} = \frac{R^{\pi,+}_{A/D}(x,z)}{B^+_{A/D}(z)} .
    \label{eq_emc_slope}
\end{equation}

The most recent global analysis~\cite{zurita2021medium} of the HERMES data reveals significant medium effects of FFs in heavy nuclei varying in $z$. On the other hand, it also suggests a small modification effect in the $^4$He nucleus, e.g., the fit shows $\sim$ 1\% (5\%) reduction at $z=0.3$ ($z=0.8$) with big error-bands. One can reasonably believe much smaller nuclear effects in the FFs of $^2$D, $^3$H, and $^3$He with loosely bound nucleons, and assume that such small nuclear effects are likely to be canceled in their ratios. We can safely treat:
\begin{equation}
   B^{\pi,\pm}_{H/T}(z) \simeq B^{\pi,\pm}_{H/D}(z) \simeq B^{\pi,\pm}_{T/D}(z) \simeq 1.\label{eq_blight},
\end{equation}
where "D", "T" and "H" denote $^2$D, $^3$H and $^3$He, respectively. One can also verify this assumption by measuring $B^+_{A/D}(z)$ via Eq.~\eqref{eq_emc_slope} where $R^A_{EMC}$ and $R^{\pi,+}_{A/D}(x,z)$ are experimental observables in DIS and SIDIS reactions.

 We can then deduce Eq.~\eqref{eq_rpm} under three different hypotheses of the EMC effect discussed in the previous section:
\begin{itemize}
\item \textbf{The EMC effect is flavor-independence and all nucleons are modified.} 

Taking $\tilde{f}(x)=R_{A}\cdot f(x)$ as the commonly modified quark-PDF in nucleus-A, we have:
\begin{equation}
    u_A =R_{A} \frac{Zu + Nd}{A},d_A = R_{A}\frac{Zd + Nu}{A}, \label{eq_npdf}
\end{equation}
with analogous expressions for the anti-quarks. From Eq.~\eqref{eq_emc_slope}, $R_A$ can be linked to the EMC ratio with the isovector correction factor, $f_{A/D}^{iso}=2(Z\sigma_{p,free}+N\sigma_{n,free})/A(\sigma_{p,free}+\sigma_{n,free})$, commonly used in the inclusive DIS measurement:
\begin{equation}
    R^A_{EMC}=R_{A}\cdot \frac{2}{A}\cdot\frac{9U+6D}{5U+5D}=R_{A}\cdot f_{A/D}^{iso}.
\end{equation} 
where $U=u+\bar{u}, D=d+\bar{d}$, Also let $u_v=u-\bar{u}, d_v=d-\bar{d}$, the ratios of $^3$H and $^3$He to $^2$D become: 
\begin{align}
    R^{\pi,-}_{H/D}&=R_H\cdot\frac{2}{3}\cdot\frac{7u_v+2d_v}{3u_v+3d_v},& 
    R^{\pi,-}_{T/D}&=R_T\cdot\frac{2}{3}\cdot\frac{7d_v+2u_v}{3u_v+3d_v},& 
    \label{eq_rpm_emc_ind}
\end{align}
In this analysis, we have used the EMC ratios from the fit to the SLAC results~\cite{Gomez94} assuming the EMC effect is isospin-independent, i.e. $R_H=R_T$. As a comparison, we also include the KP model~\cite{KULAGIN2006126,Kulagin:2010gd} which  was used in the MARATHON analysis~\cite{Adams21} to describe the different medium effects in both $^3$H and $^3$He. Although the KP model has sophisticated nuclear corrections on the structure functions of both isotopes (here $R_H \neq R_T$), it doesn't directly provide the flavor-dependent information.  Hence, the KP model is isospin-dependent but flavor-independent.

\vspace{0.1in}
\item \textbf{The EMC effect is flavor-dependent and one of the valance-quarks is more modified in an asymmetric nucleus.}

Ref~\cite{Cloet12,cloetpc} suggests that in a proton-rich nucleus (e.g. $Pb$), the $u$-quarks are more likely to be modified than $d$-quarks. Based on the isospin symmetry, it is natural to assume that $d$-quarks would have a stronger EMC effect than $u$-quarks in a neutron-rich nucleus. Despite the smaller nuclear effects that are expected in $A=3$ nuclei, they are the largest asymmetric nuclei that have opposite $Z/N$ ratios (\textit{i.e.}, mirror nuclei). It is also reasonable to assume that the medium effects between $u$ in $^3$He and $d$ in $^3$H are the same. When labeling a smaller (weaker) EMC effect as $R^+$ ($R^-$), we have:
\begin{equation}
\begin{aligned}
    u_{H} &= \frac{2\tilde{u}_p+\tilde{u}_n}{3} = \frac{2R^+ u + R^- d}{3},&
    d_{H} &= \frac{2\tilde{d}_p+\tilde{d}_n}{3} = \frac{2R^- d + R^+ u}{3},&\\
    u_{T} &= \frac{\tilde{u}_p+2\tilde{u}_n}{3} = \frac{ R^- u +2R^+ d}{3},&
    d_{T} &= \frac{\tilde{d}_p+2\tilde{d}_n}{3} = \frac{ R^+ d +2R^- u}{3}.& 
\end{aligned}
\label{eq_flavor_emc_ud}
\end{equation}

$R^{\pm}$ can be related to the EMC ratio as:
\begin{equation}
    R_{EMC}^H =\frac{2}{3}\cdot\frac{9R^+\cdot U +6R^-\cdot D}{5U+5D},~~~R_{EMC}^T =\frac{2}{3}\cdot\frac{6R^-\cdot U+9R^+\cdot D }{5U+5D} 
\end{equation}
From Eq.~\eqref{eq_rpm}, we have:
\begin{align}
\centering
    R^{\pi,-}_{H/D}&=\frac{2}{3}\cdot\frac{7R^+u_v+2R^-d_v}{3u_v+3d_v},& 
    R^{\pi,-}_{T/D}&=\frac{2}{3}\cdot\frac{2R^-u_v+7R^+d_v}{3u_v+3d_v},& \label{eq_rpm_emc_ind_asym} 
\end{align}

In the absence of existing theoretical calculations of the different nuclear effects in individual quarks in $A=3$ nuclei, one can examine two extreme cases where only $d$ ($u$) in $^3$H ($^3$H) is modified, e.g. $R^-=1$, or vice versa ($R^+=1$) assuming the real effect is opposite to the model prediction. The strengths of the modification ($R_{EMC}^{H(T)}$) match the SLAC fit or the KP model. 
\vspace{0.1in}
\item \textbf{The EMC effect is flavor-dependent and only $np$-SRC pairs are modified.}

We assume $\tilde{f}(x)=R^{np}_{A}\cdot f(x)$ as the modified quark PDF in  the nucleon from the $np$-SRC pair, and $R^{np}_A$ is the medium modification factor. There is one $np$ pair plus a "not-modified" neutron (proton) in $^3$H ($^3$He), so we can recalculate the average $u$ and $d$ quark in each isotope:
\begin{equation}
    \begin{aligned}
    u_{H} = \frac{(\tilde{u}_p+\tilde{u}_n)_{np}+u_p}{3} = \frac{R^{np}_H(u+d)+u}{3},\\
    d_{H} = \frac{(\tilde{d}_p+\tilde{d}_n)_{np}+d_p}{3} = \frac{R^{np}_H(u+d)+d}{3},\\
    u_{T} = \frac{(\tilde{u}_p+\tilde{u}_n)_{np}+u_n}{3} = \frac{R^{np}_T(u+d)+d}{3},\\
    d_{T} = \frac{(\tilde{d}_p+\tilde{d}_n)_{np}+d_n}{3} = \frac{R^{np}_T(u+d)+u}{3}. 
\end{aligned}
\label{eq_flavor_emc}
\end{equation}
We also can link $R^{np}_A$  to the EMC ratios:
\begin{equation}
    R^H_{EMC}=\frac{2}{3}(R^{np}_{H}-1)+f^{iso}_{H/D},~~~ 
    R^T_{EMC}=\frac{2}{3}(R^{np}_{T}-1)+f^{iso}_{T/D}.
\end{equation}
Plugging Eq.~\eqref{eq_flavor_emc} into Eq.~\eqref{eq_rpm}, we have:
\begin{align}
R^{\pi,-}_{H/D}(x) &=\frac{2}{3}(R^{np}_{H}+\frac{4u_v-d_v}{3u_v+3d_v}),& 
R^{\pi,-}_{T/D}(x) &=\frac{2}{3}(R^{np}_{T}+\frac{4d_v-u_v}{3u_v+3d_v}),&
 \label{eq_rpm_emc_flav_src} 
 \end{align}

The values of $R^{np}_{H(T)}$ were calculated with the strengths of the modification ($R_{EMC}^{H(T)}$) matching the SLAC fit or the KP model. 


\vspace{0.1in}
\end{itemize}



\begin{figure}[ht!]
    \centering
\includegraphics[width = 1.0\linewidth]{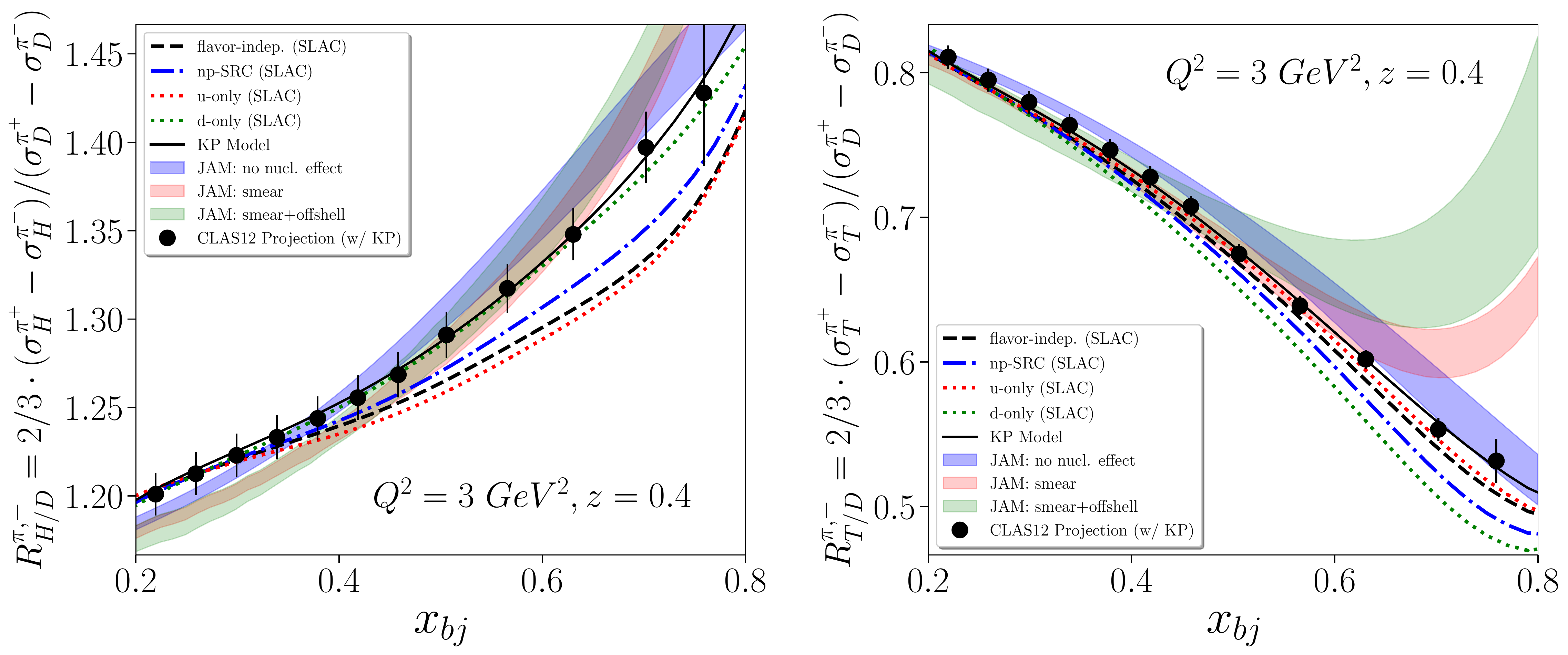}
    \caption{Projected ratios of $R^{\pi,-}_{A/D}$ for $^3$H (left) and $^3$He (right). Dashed line $R^{\pi,\pm}_{A/D}$ are calculated with the SIDIS structure functions constructed with various nuclear-PDF models discussed in the text. The data points are based on the SIDIS MC events with standard SIDIS cuts and one fixed $z$-bin ($0.35<z<0.45$). 1\% point-to-point systematic uncertainties are combined with the statistical errors.}    \label{flavor_emc}
\end{figure}

Fig.~\ref{flavor_emc} shows the experimental projection of the $R^{\pi,-}_{H/D}$ (left) and $R^{\pi,-}_{T/D}$(right) where curves represent different hypotheses discussed above. The JAM calculations are also given for comparison with the JAM PDFs from the QCD global analysis in Ref.~\cite{Cocuzza:2021rfn}. The JAM PDFs in He and T are estimated using Fermi smearing of the JAM on-shell nucleon PDFs, while allowing for off-shell nuclear effects to be flavor dependent.

Even after the tight SIDIS cuts and a single 0.35<$z$<0.45 bin is chosen, the statistical uncertainties are still small enough even at very high $x$ thanks to the wide acceptance of CLAS12. The projected data should be precise enough to distinguish small differences from these hypotheses and probe any indications of the flavor dependence of the EMC effect in $A=3$ nuclei.

Although the whole exercise above is simplified based on the LO approximation, it should provide meaningful insights into the flavor-dependent EMC effect in the $A=3$ isotope and explore any unseen isospin effects for the first time. More sophisticated global analysis, where higher-order effects, nuclear effects, and other effects are treated carefully, will be performed when the future high-precision cross-section data become valuable.

\subsection{$d/u$ Measurement}
The study of $d/u$ ratio traditionally relies on the measurement of the DIS cross-sections ($\sigma_p/\sigma_n$) which contain contributions from all quark-flavors and also suffer from large uncertainty due to the treatment of the ``effectively-free" neutrons in light nuclei.  The MARATHON experiment used the $A=3$ mirror nuclei to extract the $F_2^p/F_2^n$ ratio from inclusive DIS data with the idea that their nuclear effects, which are both small and very similar, can be largely canceled in the ratio. 
However, a recent theoretical analysis~\cite{Cocuzza:2021rfn}~\cite{Segarra:2021exb} suggests that there is still a significant residual nuclear effect between two nuclei even at high-x that can affect the $d/u$ value at x=1. The SIDIS data are essential to clarify the possible different nuclear effects in $A=3$ nuclei even at high-x~\cite{Cocuzza:2021rfn} ~\cite{Segarra:2021exb} and to cross-check the MARATHON results.

After the medium effects of $u$ and $d$ quarks in $^3$H and $^3$He are studied as discussed in the previous subsection, we can construct their own super-ratio of the pion charge-difference (ignoring $\bar{u}, \bar{d}$):
\begin{align}
    R^{\pi,-}_{H/T} = \frac{(\sigma_{H}^{\pi^+}  -  \sigma_{H}^{\pi^-})}{(\sigma_{T}^{\pi^+}  -  \sigma_{T}^{\pi^-}) }\simeq \frac{4u_{H} - d_{H}}{4u_{T} - d_{T}}, 
    \label{eq_rm_ht} 
\end{align}
which allows the direct extraction of the $d/u$ ratio from Eq~\eqref{eq_npdf_pA}:
\begin{equation}
  \frac{\tilde{d}}{\tilde{u}} \simeq \frac{7-2R^{\pi,-}_{H/T}}{7R^{\pi,-}_{H/T}-2},
 \label{eq_du_ratio}
\end{equation}
If the EMC effect is flavor-independent (Eq~\eqref{eq_npdf}), we have $ \tilde{d}/\tilde{u}=d/u $.
\begin{figure}[ht!]
    \centering
    \includegraphics[width = 0.99\linewidth]{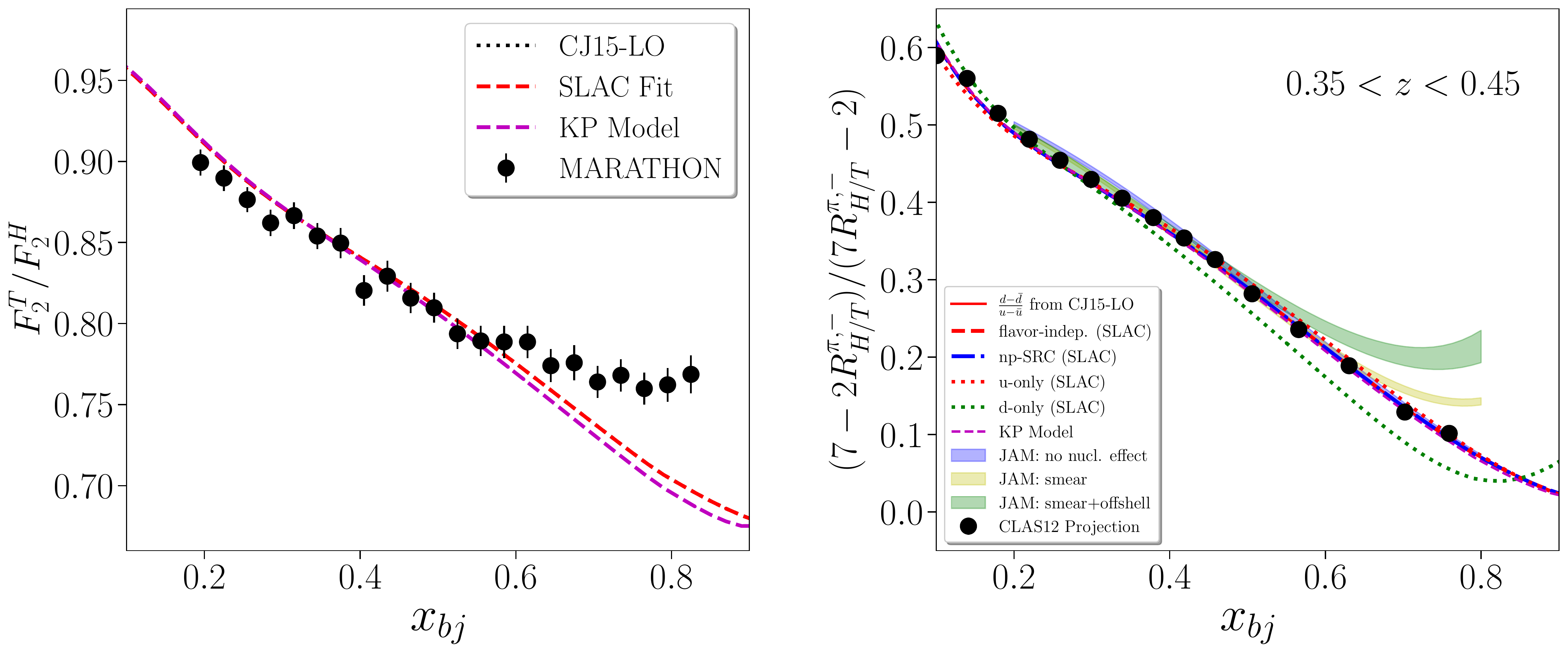}
    \caption{Projected $d/u$ ratios (left) compared with the MARATHON data (right). The mean $x$ values (statistical errors) of the highest three data points in this projection are: 0.70 (1.1\%), 0.76 (2.7\%) and 0.82 (9.5\%). Statistical errors only. Additional 1\% point-to-point systematic uncertainties are also included in the error-bars.}
    \label{du_he3_h3}
\end{figure}

Fig.~\ref{eq_rm_ht} shows the projected results based on Eq.~\eqref{eq_du_ratio} using different models discussed in the previous sections. One can see that the data points are generally very precise and can reach $x=0.82$ which matches the last data point of the MARATHON result. Note that the MARATHON results give $F^n_2/F^p_2$ ratios from the inclusive cross-sections (hence are not sensitive to flavor-dependent models), while this experiment directly probes the $d/u$ values without making nuclear corrections or assumptions. The curves suggest that the EMC effect from different models can be largely canceled in the super-ratio at low-x. However, as predicted by JAM analysis, the conventional nuclear effects (e.g., Fermi-smearing and off-shell) become problematic at $x>0.6$ and won't be totally canceled in $A=3$ nuclei. These effects can only be examined in SIDIS. If a more significant flavor-dependent EMC effect and/or nuclear effects are observed in this experiment, additional corrections can be applied to the $d/u$ ratio from the analysis discussed in the previous subsection.

\section{Conclusions}
Our new data-driven experiment will use 10.6 GeV electrons scattering off $^2$D, $^3$H and $^3$He to measure the $\pi^{\pm}$ SIDIS events using the standard configuration CLAS12 detector along with a new tritium target system designed by the approved SRC experiment (E12-20-005). This unique opportunity allows us to use $A=3$ nuclei to investigate the nature of the EMC effect and the $d/u$ ratio at high-x. We aim to perform 4D binning of the SIDIS data in all three nuclei, thanks to the usage of the mirror isotopes with well-controlled nuclear effects, plus the wide kinematic coverage and high count rates of CLAS12. The high-quality data will be crucial to the TMD community to test QCD factorization and to study hadronization in SIDIS. Fully mapping out the $P_T$ distribution in the $A=3$ SIDIS data will help to decouple the unpolarized TMDs and FFs of $u$ and $d$ quarks and reveal, for the first time, their nuclear effects. Such a measurement is in time for the TMD program during the JLab 12GeV-era and before the future EIC.

\vspace{6pt} 
\acknowledgments{Z.H. Ye is supported by Tsinghua University Initiative Scientific Research Program.}






\bibliography{tritium.bib}

\end{document}